\documentclass{PoS}
\usepackage{amsmath,amssymb,latexsym, cancel}
\usepackage{amssymb,latexsym}
\usepackage{graphicx}
\usepackage{float}

\title{Free Energy of a Large-$N$ Pion Gas and Chiral Symmetry Restoration}

\ShortTitle{Free Energy of a Large-$N$ Pion Gas and Chiral Symmetry Restoration}

\author{\speaker{Santiago Cort\'es} \\
        Departamento de F\'{\i}sica, Univ. de Los  Andes, 111711 Bogot\'a, Colombia.\\
        E-mail: \email{js.cortes125@uniandes.edu.co}}

\author{\'Angel G\'omez Nicola\\
        Departamento de F\'{\i}sica Te\'orica and UPARCOS. Univ. Complutense. 28040 Madrid. Spain. \\
        E-mail: \email{gomez@ucm.es}}
				
\author{John Morales\\
        Departamento de F\'{\i}sica, Univ. Nacional de Colombia, 111321 Bogot\'a, Colombia.\\
        E-mail: \email{jmoralesa@unal.edu.co}}

\abstract{We study thermal properties of a large-$N$ massless pion gas using a low-energy QCD approach given by an $O(N+1)/O(N)$ Nonlinear Sigma Model. We build diagrammatically the associated finite free energy to $\mathcal{O}(TM^{3})$ in the pion mass expansion through an effective vertex that considers all the contributions coming from closed diagrams. Subsequently, we calculate finite order parameters such as the quark condensate and its respective derivative, i.e., the scalar susceptibility, in the chiral limit, along with their associated critical exponents. These results are compared with our previous unitarized scattering analyses for the chiral transition universality class, thus showing a reasonable agreement both with lattice simulations and these resonant studies.}

\FullConference{XVII International Conference on Hadron Spectroscopy and Structure - Hadron2017\\
		25-29 September, 2017\\
		University of Salamanca, Salamanca, Spain}

\begin{document}

\section{Introduction and Formalism}

\begin{figure}
\centering
\includegraphics[scale=0.9]{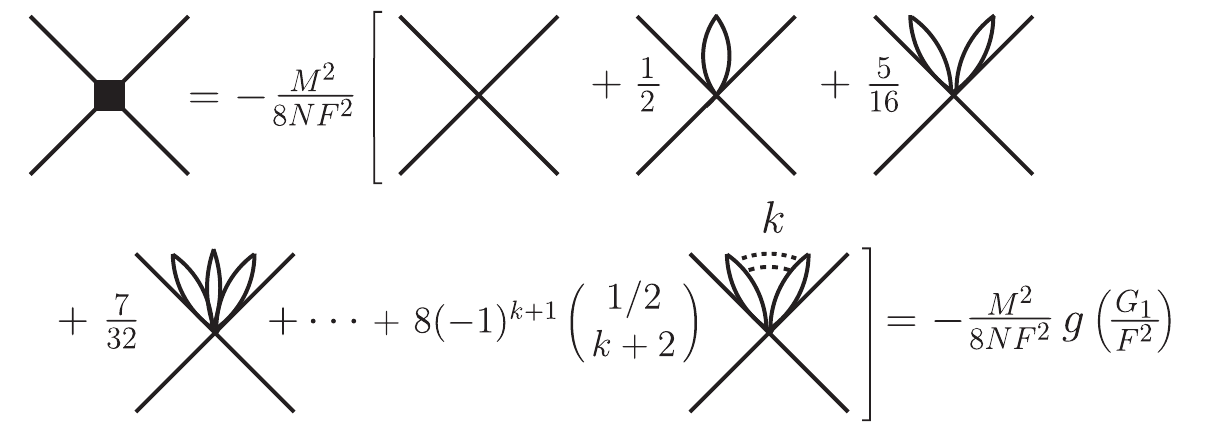}
\caption{Massive effective thermal vertex.}
\label{fig:massvertex}
\end{figure}

Low-energy QCD phenomena such as chiral symmetry restoration are essential to understand the behavior of hadron matter created in heavy-ion collision experiments such as LHC-ALICE and RHIC, a fact showed and confirmed by lattice simulations \cite{Aoki:2009sc}. Our purpose is to study the critical behavior of a Large-$N$ pion gas through its free energy by considering an $O(N+1)/O(N)$ Nonlinear Sigma Model with an explicit chiral symmetry breaking term given by the following Lagrangian \cite{Dobado:1994fd,dobadobook}:

\begin{align}
\mathcal{L}_{NLSM}&=\frac{1}{2}\left[\delta_{ab}+\frac{1}{NF^{2}}\frac{\pi_{a}\pi_{b}}{1-\pi^2/NF^{2}}\right]\partial_{\mu}\pi^{a}\partial^{\mu}\pi^{b}+NF^2M^2\sqrt{1-\frac{\pi^2}{NF^2}}, \label{NLSM} \\
\pi^2&=\displaystyle\sum_{a=1}^{N}\pi_{a}\pi^{a}, \notag
\end{align}

\noindent where $M^2$ and $\sqrt{N}F$ are respectively the pion mass and  the pion decay constant in the chiral limit. Large-$N$ closed dominant diagrams ($\mathcal{O}(N)$) are obtained from the mass vertex in (\ref{NLSM}). Momentum contributions are subdominant either by their large-$N$ contribution (i.e., $\mathcal{O}(N^{0})$) or their associated Feynman rules are canceled due to their odd parity. Bearing this in mind, we proceed to build a massive effective thermal vertex as it is shown in Figure \ref{fig:massvertex}.

This vertex takes into account all possible insertions of thermal tadpoles coming from diagrams with six or more external legs. Its Feynman rule is written in terms of a function $g(x)$ that reads

\begin{equation}
g(x)=-\frac{8}{x^2}\left[\sqrt{1-x}-1+\frac{x}{2}\right]=-8\sum_{k=0}^{\infty}(-1)^k \begin{pmatrix} 1/2\cr k+2 \end{pmatrix} x^k=1+\frac{x}{2}+\frac{5}{16}x^2+\cdots
\label{eqn:massvert}
\end{equation}

\noindent $G_{1}(M,T)$ is taken as in \cite{Gerber:1988tt}. Since this is not a scattering approach, we do not consider any combinatory factor linked with the way the pion lines are attached either with external legs or loops. Now, we are able to diagramatically construct the partition function for $N$ massless pions.


\section{Free Energy and Order Parameters}

\subsection{Free Energy}

\begin{figure}
\centering
\includegraphics[scale=0.55]{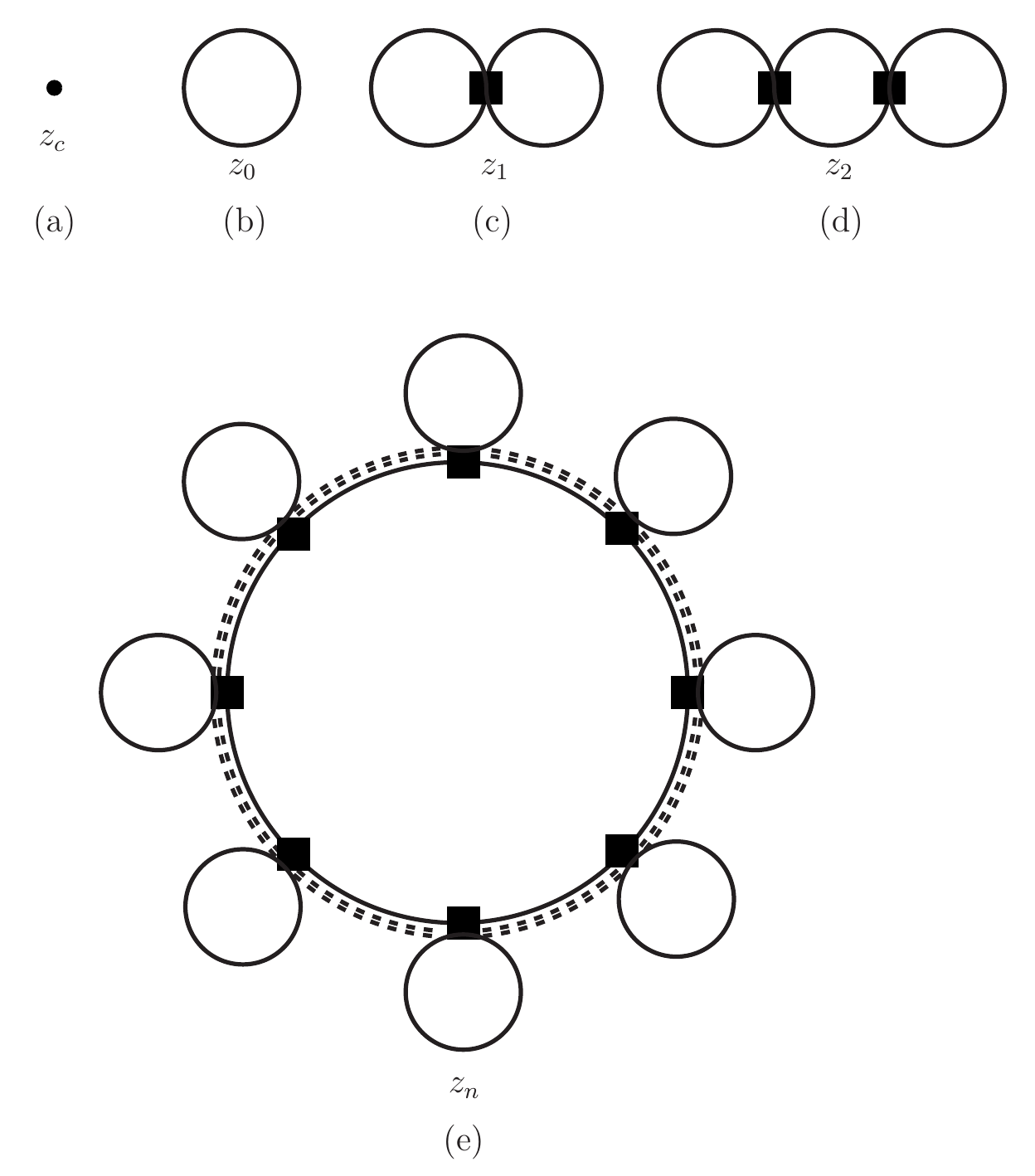}
\caption{Nondivergent Feynman diagrams of a large-$N$ free energy.}
\label{fig:partfun}
\end{figure}

By taking the limit $M/F\rightarrow 0$ and considering large-$N$ dominant contributions up to $\mathcal{O}(M^{3}T)$, we obtain a free energy that does not need to be renormalized after examining the contributions coming from the closed diagrams showed in Figure \ref{fig:partfun} \cite{Cortes:2016ecy}. The dot (a) represents the constant term that comes from the expansion of (\ref{NLSM}), whereas the Daisy diagram (e) depicts multiple insertions of the mass vertex along a single loop. Any other diagram not listed in the figure is either large-$N$ subdominant or contributes as $\mathcal{O}(M^{4})$ (e.g., foam-like diagrams), which are divergent terms avoided in this limit. The explicit form of the free energy is showed in equation (\ref{eqn:parfun}), whose corresponding Feynman diagrams are listed in Figure \ref{fig:partfun}.

\vspace{-0.5cm}

\begin{align}
 z(M,T)&=-N\frac{\pi^{2}T^{4}}{90} -NM^2F^2\left\{ 1-\frac{T^{2}}{24F^{2}}+h\left(\frac{T^2}{12F^2}\right)\right\} -\frac{NM^3T}{8\pi}\left\{\frac{2}{3}-2h'\left(\frac{T^{2}}{12F^{2}}\right)\right. \notag \\
&\left. +H\left[-\frac{1}{2} \tilde g\left(\frac{T^2}{12F^2}\right)\ \right]\right\}. \label{eqn:parfun}\end{align}

The corresponding functions $h(x)$ and $H(x)$ read
\begin{align}
 h(x)&=\sqrt{1-x}-1+\frac{x}{2},\text{ }\tilde g(x)=\frac{1}{\sqrt{1-x}}-1, \\
 H(x)&=x^2+2\sum_{n=3}^\infty  \frac{(2n-5)!!}{n!} x^n=-\frac{2}{3}\left(1-3x-\sqrt{1-2x}+2x\sqrt{1-2x}\right). 
\end{align}

\subsection{Scalar Quark Condensate and Susceptibility}

\begin{figure}
\begin{center}
\centerline{\includegraphics[width=.53\linewidth]{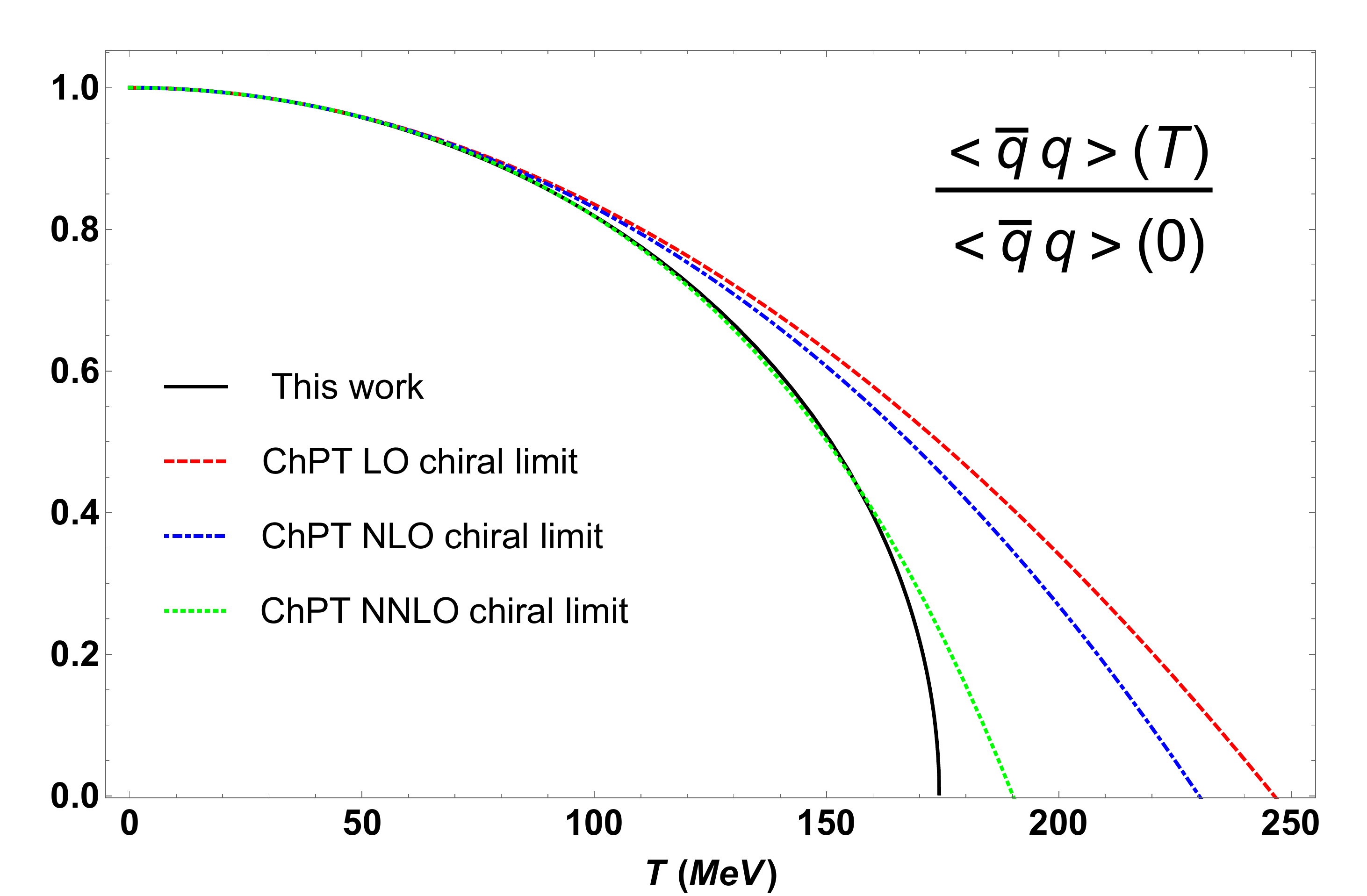} \includegraphics[width=.53\linewidth]{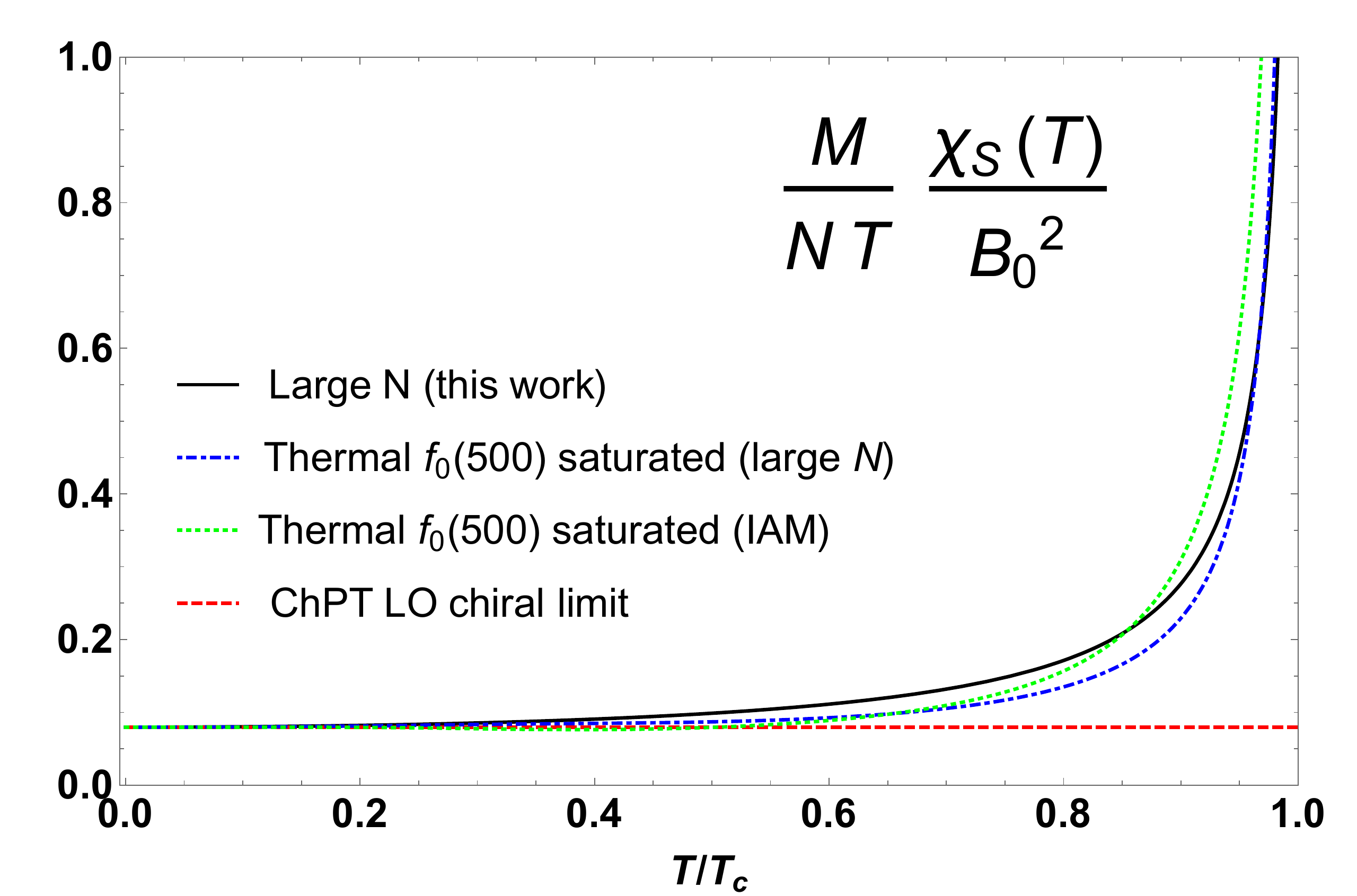}}
\end{center}
\caption{Normalized scalar quark condensate and scalar susceptibility.}
\label{fig:scalarcond}
\end{figure}

First and second derivatives of this partition function allows us to respectively find the quark condensate and the scalar susceptibility, whose thermal evolution in the chiral limit $M\rightarrow 0^{+}$ is plotted in Figure \ref{fig:scalarcond} \cite{Cortes:2016ecy}.

We also plot the $f_{0}(500)$-saturated scalar susceptibility \cite{Cortes:2015emo}, along with the well-known ChPT results \cite{Gerber:1988tt}. We find in an exact way that the quark condensate scales as $(1-T^{2}/T_{c}^{\,2})^{\beta}$ with $\beta=\text{0.5}$ in the critical region, whilst the critical exponent of the susceptibility is $\gamma=\text{0.75}$, where $\chi_{S}(T)\sim (1-T^{2}/T_{c}^{\,2})^{-\gamma}$. Furthermore, we check that the condensate cancels out exactly at a critical temperature $T_{c}=\sqrt{12}F\approx\text{174 MeV}$, thus showing that the system undergoes a second-order phase transition. 

\section{Conclusions} 

Our results show that this critical behavior rationally agrees with those given for the universality classes expected from lattice simulations. However, our numerical values obtained for $\gamma$ and $T_{c}$ differ from our previous scattering analysis since this approach only takes into account contributions coming from Nambu-Goldstone bosons.

\section*{Acknowlegdments}

SC wants to thank Facultad de Ciencias and Vicerrector\'{\i}a de Investigaciones of Universidad de los Andes for financial support.


\begin{thebibliography}{99}
\bibitem{Aoki:2009sc}
  Y.~Aoki, S.~Borsanyi, S.~Durr, Z.~Fodor, S.~D.~Katz, S.~Krieg and K.~K.~Szabo,
  JHEP {\bf 0906}, 088 (2009).
	
\bibitem{Dobado:1994fd} 
  A.~Dobado and J.~Morales,
  Phys.\ Rev.\ D {\bf 52}, 2878 (1995).
	
\bibitem{dobadobook}
A. Dobado, A. G\'omez Nicola, A. L. Maroto, J. R, Pel\'aez, \emph{Effective Lagrangians for the Standard Model}. Springer-Verlag Berlin Heidelberg (1997)
	
	
\bibitem{Gerber:1988tt}
  P.~Gerber and H.~Leutwyler,
  Nucl.\ Phys.\  B {\bf 321}, 387 (1989).
	
	
\bibitem{Cortes:2016ecy} 
  S.~Cort\'es, A.~G\'omez Nicola and J.~Morales,
  Phys.\ Rev.\ D {\bf 94}, no. 11, 116008 (2016).
	

\bibitem{Cortes:2015emo} 
  S.~Cort\'es, A.~G\'omez Nicola and J.~Morales,
  Phys.\ Rev.\ D {\bf 93}, no. 3, 036001 (2016).


\end{thebibliography}
\end{document}